# Data-driven Optimal Power Flow: A Physics-Informed Machine Learning Approach

Xingyu Lei, *Student Member*, *IEEE*, Zhifang Yang, *Member*, *IEEE*, Juan Yu, *Senior Member*, *IEEE*, Junbo Zhao, *Senior Member*, *IEEE*, Qian Gao, *Student Member*, *IEEE*, Hongxin Yu

***Abstract*—This paper proposes a data-driven approach for optimal power flow (OPF) based on the stacked extreme learning machine (SELM) framework. SELM has a fast training speed and does not require the time-consuming parameter tuning process compared with the deep learning algorithms. However, the direct application of SELM for OPF is not tractable due to the complicated relationship between the system operating status and the OPF solutions. To this end, a data-driven OPF regression framework is developed that decomposes the OPF model features into three stages. This not only reduces the learning complexity but also helps correct the learning bias. A sample pre-classification strategy based on active constraint identification is also developed to achieve enhanced feature attractions. Numerical results carried out on IEEE and Polish benchmark systems demonstrate that the proposed method outperforms other alternatives. It is also shown that the proposed method can be easily extended to address different test systems by adjusting only a few hyperparameters.**

***Index Terms*—Feature decomposition, multi-parametric programming (MPP), network architecture, optimal power flow, sample classification, stacked extreme learning**

## I. INTRODUCTION

OPTIMAL power flow (OPF) is one of the most important tools for power system analysis, such as market clearing, network optimization, voltage control, generation dispatch, etc. However, the nonlinearity and nonconvexity of the OPF model lead to a high computational burden. Indeed, when a high proportion of renewable energy is connected to the grid, the OPF calculation needs to be solved in real-time to determine the power system optimal operation strategy. The power flow linearization approaches may partially simplify the complexity [1], [2], but may lead to local optimal solution [3], [4]. Although several signs of progress have been made, the computational efficiency is still a bottleneck. For example, the OPF problem needs to be iteratively solved for an enormous number of samples in the probabilistic analysis, which has been a crucial tool to identify the hidden risk in power systems with high uncertainties [5]. Some studies relieve the computational burden of OPF calculation by reducing the modeling complexity at the expense of accuracy sacrifice [6]-[8]. Also, there are many heuristic methods proposed to solve OPF problems and they have been shown to achieve desired robustness and convergence for solving complex OPF problem with nonconvex cost functions, discrete control variables, and prohibited unit operating zones [9]-[11]. The model-based OPF calculation still needs numerous iterations inside the algorithm, which may lead to a high computational burden for large-scale systems [12], [13]. By comparison, our work addresses this problem by developing a data-driven approach that treats the OPF problem as a functional mapping between the system operating status and the OPF solutions.

In recent years, the data-driven method has been widely applied in power system analysis, including the estimation of distribution factors [14], the Jacobian matrix [15] and the admittance matrix from noisy synchrophasor data [16], the suppression of uncertainties [17]-[20] and regression [21]-[22]. In particular, neural networks (NNs) have been widely used in power systems [23]-[27]. One of the challenges for NNs is that a large number of hyperparameters need to be adjusted artificially. The performance of an NN-based algorithm depends mostly on the selection of its hyperparameters, but there is no efficient hyperparameters adjustment algorithm to guide this process.

Compared with the traditional NN-based algorithms [28], [29], the stacked extreme learning machine(SELM) is a novel machine learning technology that randomly generates the input weights of hidden layer neurons and analytically determines the output weights through simple matrix computations. This significantly improves the training speed while requiring fewer hyperparameters to be tuned [30], [31]. Meanwhile, it splits a sizeable neural network into several serially computed smaller ones to achieve less memory occupation and higher feature extraction capability [32]. In practice, SELM has been used for both regression and classification [32]-[35]. However, SELM has a limited learning ability because of the random generation of input weights and the analytical output weight determination process. Further adjustment of the SELM is required to accurately learn the features of the OPF problem.

The objective of the OPF problem is to obtain solutions according to the system operating status while respecting various constraints. If the complicated relationship between the system operating status and the OPF solutions can be learned by machine learning technology, its computational efficiency will be significantly improved. Note that an efficient data-driven OPF analysis algorithm needs not only high precision

This work was supported in part by Natural Science Foundation of China (No. 51807014) and Science and Technology Project of State Grid Corporation of China (5100-201999333A-0-0-00).

X. Lei, Z. Yang, J. Yu, and Q. Gao are with State Key Laboratory of Power Transmission Equipment & System Security and New Technology, Chongqing University, Chongqing, 400030 China (email: lxylxy7@163.com; yangzfang@126.com; 148454745@qq.com; 875366843@qq.com).

J. Zhao is with the Department of Electrical and Computer Engineering, Mississippi State University, Starkville, MS 39762 USA (e-mail: junbo@ece.msstate.edu).

H. Yu is with State Grid Chongqing Electric Power Company Electric Power Research Institute, Chongqing, 401123 China.

and fast computing speed, but also good generalization capability, which makes the unique characteristics of SELM (e.g. fast training, less intervention, and small memory occupation) an ideal candidate. However, the relationship between the input (system operating status) and the output (the optimal power flow solutions) is rather complex. Hence, direct learning using the original SELM is intractable which will be shown in the simulation results. Fortunately, the physical model of OPF is known and this motivates us to develop a new framework to reduce the learning complexity of SELM by including its physical characteristics.

To this end, a physics-informed data-driven OPF approach is proposed. Compared with the current model-based ones, it has the following advantages: 1) the time-consuming iterations of OPF calculation are avoided, which are replaced by the direct SELM mapping; 2) the system topologies and parameters are not required; and 3) a high-quality solution can be obtained in a short time, which may provide guidance information for the model-based OPF to accelerate computing speed.

The main contributions of this paper are summarized as follows:

1) A SELM learning framework is proposed for the OPF considering the physical characteristics of the OPF model. Specifically, the complex OPF model features are decomposed into three stages to correct the learning bias. Taking advantage of SELM, the massive adjustment of hyperparameters is avoided, which is the key challenge for the deep learning method. Thanks to that, the proposed approach can be easily extended to different systems with different scales. To further enhance the learning ability of each stage, a reinforcement mode is used in the hidden layer when designing the SELM network.

2) A sample pre-classification strategy based on active constraint identification is proposed to extract more effectively the features while reducing the learning complexity. Indeed, according to the *multi-parametric programming* (MPP) theory, OPF may have different features under different combinations of active constraints, yielding a highly complex problem.

The rest of this paper is organized as follows. Section II provides a data-driven OPF regression framework based on SELM, and Section III presents a sample pre-classification method to improve the learning performance followed by the summarization of the proposed approach. Section IV discusses the experimental results. The conclusion is given in Section V.

## II. A Framework of Data-Driven OPF based on SELM

This section first briefly introduces the SELM and the data-driven OPF learning framework. Then, the SELM network for learning the OPF model features is proposed.

### A. A Brief Introduction of SELM

ELM is a *single-hidden-layer feedforward network* (SLFN). The hidden layer output vector $\boldsymbol{h}(\boldsymbol{x}_{ns})$ can be expressed as:

$$\boldsymbol{h}(\boldsymbol{x}_{ns}) = \mathrm{g}(\mathbf{W}\cdot\boldsymbol{x}_{ns} + \boldsymbol{b}) \tag{1}$$

where $\boldsymbol{x}_{ns}$ is the input feature vector of $ns^{\text{th}}$ sample; $\mathrm{g}(\cdot)$ represents the activation function; The input weight matrix $\mathbf{W}$ and bias vector $\boldsymbol{b}$ are randomly generated. Then, the hidden layer output matrix $\mathbf{H}$ can be obtained by gathering $\boldsymbol{h}(\boldsymbol{x}_{ns})$ of all samples:

$$\mathbf{H}=[\boldsymbol{h}(\boldsymbol{x}_1)^T \cdots \boldsymbol{h}(\boldsymbol{x}_{Ns})^T]^T \tag{2}$$

where $Ns$ is the total number of samples.

The key idea of ELM is to calculate the output weight matrix $\mathbf{\Psi}$ between the hidden layer and the output layer in SLFN via the following equation [30]:

$$\mathbf{\Psi} = (\mathbf{H}^T\mathbf{H})^{-1}\mathbf{T} \tag{3}$$

where $\mathbf{T}$ denotes the target matrix that will be learned by ELM, which is formed in a similar way to the hidden layer output matrix $\mathbf{H}$ by gathering the target vectors of all samples.

Motivated by deep-learning models, SELM is proposed via a stacked ELM with a multilayer NN structure [15]. In order to extract the vital information from the training data, the PCA dimension reduction method is introduced to partition a large ELM NN into multiple stacked small ELMs. The first layer of SELM is an original ELM that generates the parameters of hidden layer neurons randomly. For other layers, only partial parameters are generated in a random way as some parameters are obtained from the parameters of the previous layer after dimension reduction. The information of the input data is propagated to the next layer and the input information is transmitted from layer to layer until the last one. Specifically, the output weight matrix of the $i^{\text{th}}$ iteration is denoted as $\mathbf{\Psi}^{(i)}$, which can be obtained by solving the following optimization problem with L2 regularization [31]:

$$\min_{\mathbf{\Psi}^{(i)}} \left\{ f^{(i)} = \left\|\mathbf{\Psi}^{(i)}\right\| + C\left\|\mathbf{T}-\mathbf{H}^{(i)}\mathbf{\Psi}^{(i)}\right\|_2^2 \right\} \tag{4}$$

where $\mathbf{H}^{(i)}$ is the hidden layer output matrix of the $i^{\text{th}}$ iteration, and $C$ is a penalty factor, leading to a tradeoff between the training error and the norm of output weights.

The matrix $\mathbf{\Psi}^{(i)}$ is obtained by solving $\partial f^{(i)}/\partial\mathbf{\Psi}^{(i)} = 0$, which can be expressed as follows:

$$\mathbf{\Psi}^{(i)} = \left(\frac{\mathbf{I}}{C} + \mathbf{H}^{(i)T}\mathbf{H}^{(i)}\right)^{-1}\mathbf{H}^{(i)T}\mathbf{T} \tag{5}$$

Note that there may be redundant information in the output matrix $\mathbf{H}^{(i)}$ of the $i^{\text{th}}$ iteration. Hence, the dimension of $\mathbf{\Psi}^{(i)}$ can be reduced from $L$ to $l$, where $L$ and $l$ are the original and the reduced number of the hidden neurons, respectively. In the procedure of PCA-based dimension reduction, the eigenvectors matrix $\mathbf{V}^{(i)} \in \mathbf{R}^{L\times L}$ is generated and the top $l$ eigenvectors are recorded as $\tilde{\mathbf{V}}^{(i)} \in \mathbf{R}^{L\times l}$. The original $L$ random hidden neurons can now be substituted by $l$ significant neurons, and the reduced hidden layer output matrix can be expressed as follow:

$$\tilde{\mathbf{H}}^{(i)} = \mathbf{H}^{(i)}\tilde{\mathbf{V}}^{(i)} \tag{6}$$

When the number of hidden neurons is reduced to $l$, only $L-l$ hidden neurons in the next iteration need to be generated randomly, and a new $\mathbf{H}_{new}$ can be calculated. Then, the hidden layer output matrix of the next iteration can be formulated as [32]:

$$\mathbf{H}^{(i+1)} = [\tilde{\mathbf{H}}^{(i)}, \mathbf{H}_{new}] \tag{7}$$

The output weight vector of this iteration can be calculated by (5), and further used to obtain the reduced eigenvectors $\tilde{\mathbf{V}}^{(i+1)}$ in the same way. Conducting the iteration until the dimension reduction procedure is not needed. More details about the SELM can also be found in [31], [32].

### B. Proposed Learning Framework of Data-Driven OPF

In this subsection, a data-driven OPF learning framework is proposed, which is based on a decomposition of OPF model features and an error correction process.

Model-based OPF problem is a nonlinear and nonconvex programming problem, which can be expressed as follows:

$$\min F = \sum_{i \in S_G} \left( a_{2i} PG_i^2 + a_{1i} PG_i + a_{0i} \right) \quad (8)$$

$$\left.\begin{aligned} PG_i - PD_i &= V_i \sum_{j \in i} V_j (G_{ij} \cos\theta_{ij} + B_{ij} \sin\theta_{ij}) \\ QG_i - QD_i &= U_i \sum_{j \in i} U_j (G_{ij} \sin\theta_{ij} - B_{ij} \cos\theta_{ij}) \end{aligned}\right\} \quad (i \in S_B) \quad (9)$$

$$PF_k = PF_{ij} = V_i V_j (G_{ij} \cos\theta_{ij} + B_{ij} \sin\theta_{ij}) - V_i^2 G_{ij} \quad (10)$$

$$\left.\begin{aligned} \underline{PG_i} \le PG_i \le \overline{PG_i} \\ \underline{QG_i} \le QG_i \le \overline{QG_i} \end{aligned}\right\} \quad (i \in S_G) \quad (11)$$

$$\underline{V_i} \le V_i \le \overline{V_i} \quad (i \in S_B) \quad (12)$$

$$-\overline{PF_k} \le PF_k \le \overline{PF_k} \quad (k \in S_K) \quad (13)$$

where $PD_i, QD_i$ represent the active and reactive power demand, respectively; $PF_k, QF_k, V_i, \theta_i$ are state variables that represent the active branch flow, reactive branch flow, voltage magnitude, and voltage angle, respectively; $PG_i, QG_i$ are control variables that represent the active power and reactive power output of generators, respectively; $F$ is the objective function that represents system operating cost at the optimal steady-state; $S_G, S_B, S_K$ denote the sets of power generation, the system buses, and system branches; $i, j$ are the bus indices, and $k$ represents the branch index; $a_{2i}, a_{1i}, a_{0i}$ are the generation cost coefficients, respectively; $\theta_{ij} = \theta_i - \theta_j$ is the difference of angle between the $i^{th}$ and $j^{th}$ bus; $G_{ij}, B_{ij}$ denote the conductance and susceptance of admittance between the $i^{th}$ and $j^{th}$ bus, respectively. It should be noted that renewables are treated as negative loads, which are included in $PD_i, QD_i$.

The OPF model (8)-(13) contains power system physical information, including the power network topology, the branch parameters, and the corresponding physical laws, etc. However, because of the nonlinear and nonconvex features of the model, the OPF calculation requires multiple iterations, which is time-consuming for large-scale systems. From a data-driven point of view, the OPF calculation process can be regarded as a nonlinear projection: the system power demand $PD_i, QD_i$ are taken as the input, and the OPF calculation results $PG_i, QG_i, V_i, \theta_i, PF_k, QF_k, F$ are treated as the output. The mapping relationship between the input and output can be learned off-line via historical data or simulated data.

In this paper, a data-driven OPF learning framework is proposed. It has three stages, as shown in Fig. 1. Note that the SELM is a novel computing paradigm with fast training speed and less intervention. However, the learning capacity of SELM is limited, in particular, for the problem that has a complicated mapping relationship between the input and output, such as the OPF. In this scenario, the direct learning is intractable. To relieve the learning pressure of SELM, a data-driven OPF learning framework is developed that divides the learning task into three stages. They are explained as follows:

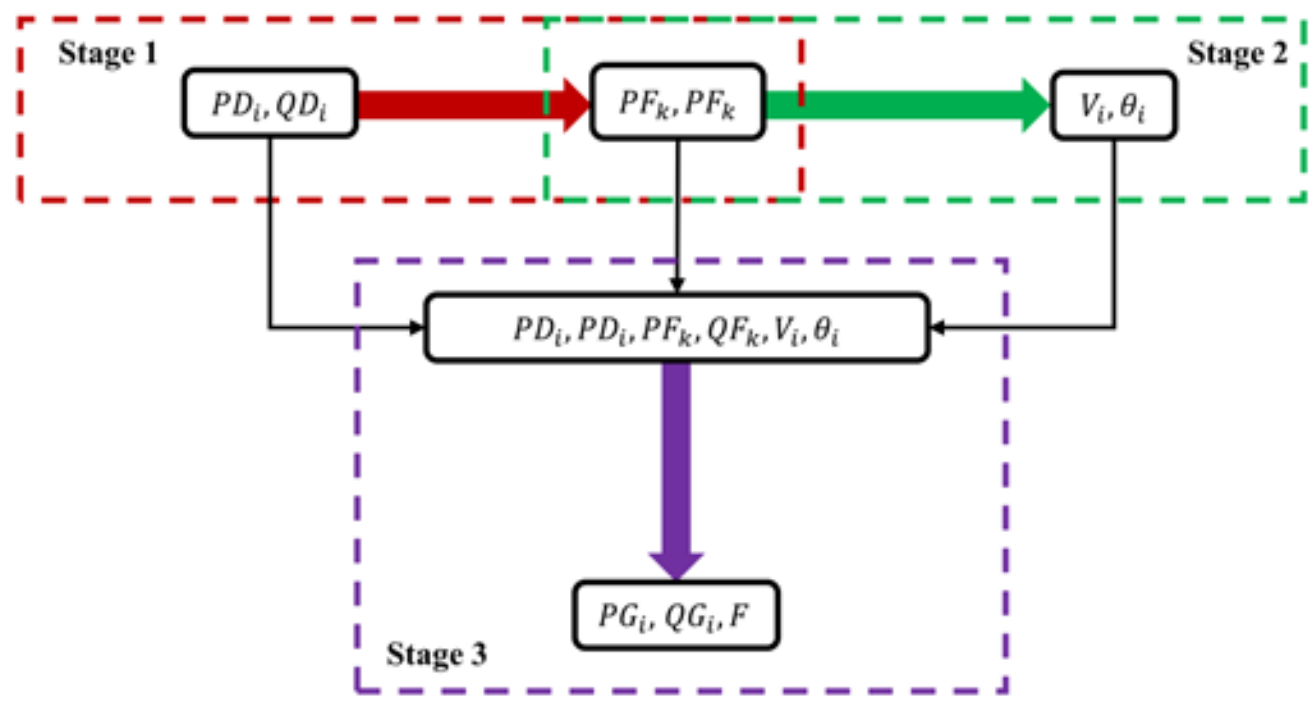

Fig. 1. Data-driven OPF learning framework.

Stage 1 ($f: PD_i, QD_i \rightarrow PF_k, QF_k$) learns the active branch flow and the reactive branch flow[1]. Through the historical power flow data $PD_i, QD_i$ and $PF_k, QF_k$, the branch flow can be decomposed from the complicated OPF model and used for SELM learning.

[1]Here we provide an example for the construction of the SELM. The power demands $PD_i$ and $QD_i$ of all buses for each sample form the input feature vector $\boldsymbol{x}_{ns}$. Then, the hidden layer output matrix $\mathbf{H}$ can be computed according to (1), (2). The target vector $\boldsymbol{t}_{ns}$ can be obtained in a similar way by gathering the branch flow $PF_k$ and $QF_k$. The target matrix $\mathbf{T}$ can be expressed as $\mathbf{T} = [\boldsymbol{t}_1^T \cdots \boldsymbol{t}_{Ns}^T]^T$. Then, the output weight matrix $\boldsymbol{\Psi}$ can be calculated by (4)-(7).

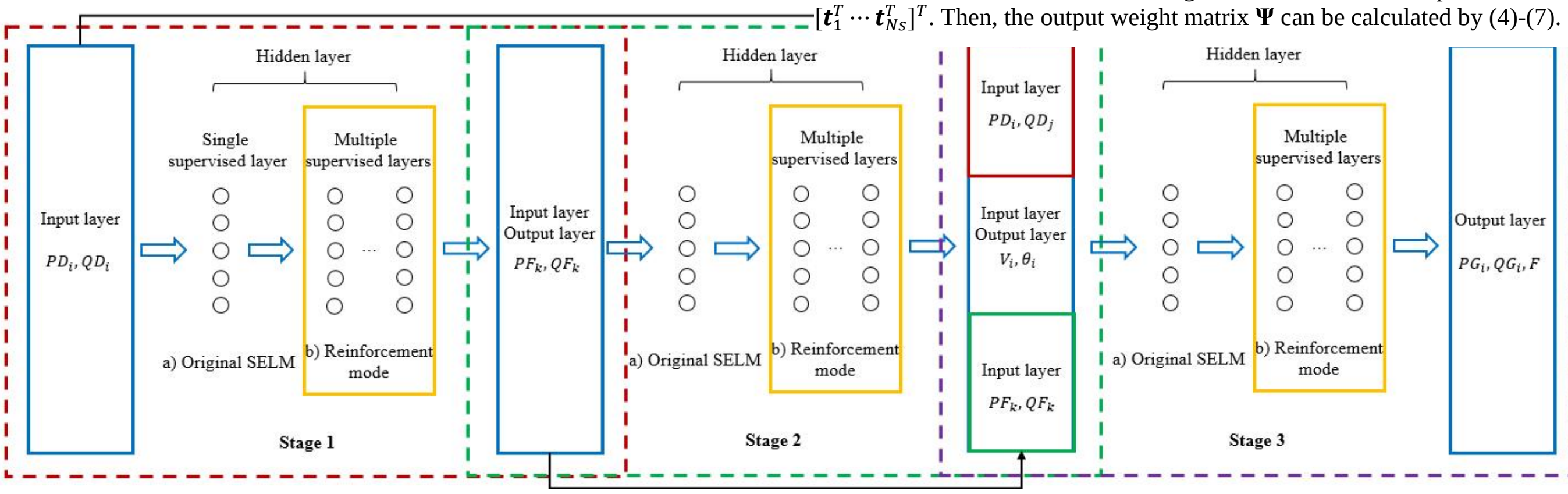

Fig. 2. Architecture of the proposed three-stage SELM network with reinforcement mode.

Stage 2 ($f: PF_k, QF_k \rightarrow V_i, \theta_i$) learns the voltage magnitude and the voltage angle, which covers the physical information in the power flow model, including the line parameter information, the corresponding physical laws, etc.

Stage 3 ($f: PD_i, QD_i, PF_k, QF_k, V_i, \theta_i \rightarrow PG_i, QG_i, F$) learns the control variables and the objective function value, which can be seen as an error correction process.

The key idea of the proposed framework is to reduce the learning difficulty of the OPF. In fact, the learning target of Stage 2 (i.e., $V_i, \theta_i$) and Stage 3 (i.e., $PG_i, QG_i, F$) can be directly calculated by the physical power flow model based on the state variables obtained from Stage 1 (i.e., $PF_k, QF_k$). Instead, Stage 2 and Stage 3 act as an error correction process in the proposed framework. Although the learning accuracy of a single SELM model in each stage may not be sufficiently high, the learning error gradually reduces through the three stages, eventually meeting the accuracy requirement. The idea of the whole framework is motivated by the ResNet, which makes a shortcut pathway directly connecting the input and the output in a middle layer [36].

## C. *Architecture Design of SELM Network*

To improve the learning accuracy, a three-stage enhanced SELM network architecture is designed and shown in Fig. 2. The input and output layer of three stages are designed to separate the learning target, thereby correcting the learning error of the OPF model. The hidden layer develops a reinforcement mode to improve the learning ability of SELM.

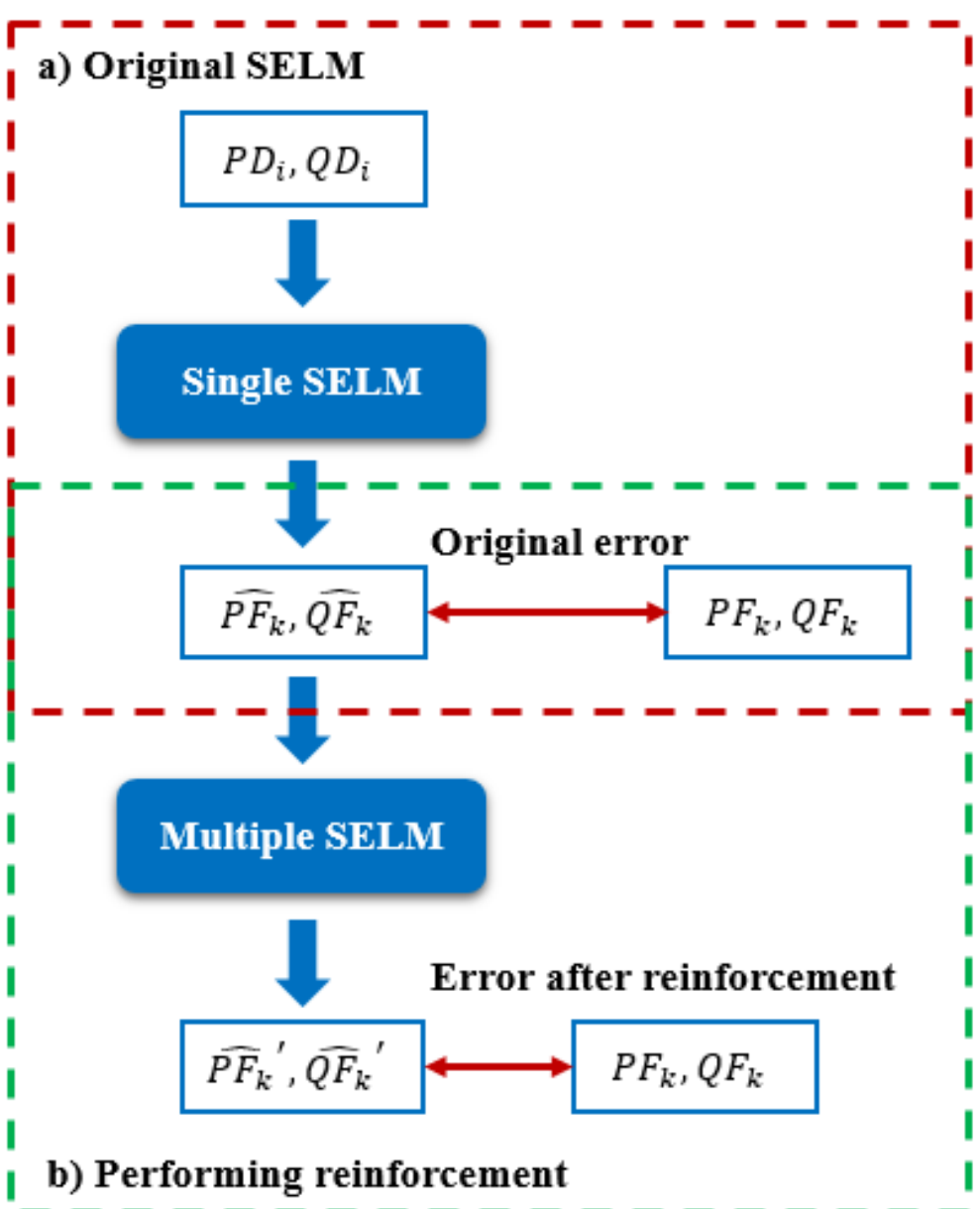

Fig. 3. Sketch of the reinforcement process in Stage 1.

For the hidden layers, the structures of the three stages are similar. In addition to the original SLEM, a reinforcement process constructed by multiple supervised layers is proposed. Taking the subnetwork of Stage 1 as an example, as shown in Fig. 3. The hidden layer of SELM includes two components:

a) A single supervised layer of the SELM regression;

b) Multiple supervised layers of the SELM regression.

The reinforcement process is constructed because of the output $\widehat{PF}_k, \widehat{QF}_k$ obtained from the input $PD_i, QD_i$ flowing through a) may deviate far from the real value $PF_k, QF_k$ in the realistic scenarios. Therefore, a reinforcement mode is used to decrease the error between $PF_k, QF_k$ and $\widehat{PF}_k, \widehat{QF}_k$ caused by single SELM learning. Note that the design for the number of layers in the reinforcement is required, aiming to achieve a trade-off between the learning performance and network complexity. Details will be described in case studies.

The above network architecture design constitutes a three-stage enhanced SELM network that can successfully reduce the learning complexity based on the OPF learning framework proposed in Section II-B and further strengthen the learning ability of SELM through the reinforcement.

# III. Sample Pre-Classification Strategy

In this section, based on the OPF model, a sample pre-classification method is proposed to achieve better learning performance.

## A. *Basic Idea of the Sample Pre-Classification*

For the OPF problem, the nonlinearity of the mapping relationship between the input and output is complex. In fact, there is a segmentation in the nonlinearity, which is relied on the active constraints of the OPF model. As illustrated in Fig. 4, based on the *multi-parametric programming* theory, the mapping relationship can be divided into several critical regions. A critical region (CR) Θ is defined as a set of inputs whose corresponding outputs share the same active constraints [37]. There is a fixed nonlinear relationship between the input and output in a specific critical region. From a data-driven perspective, if the samples can be pre-classified according to their active constraints, the nonlinear mapping relationship will be further simplified. Hence, the key idea of the proposed sample pre-classification strategy is to cluster samples with the same or similar active constraints to mitigate the complicated mapping relationship between the variation of the optimal solutions with respect to the random power demand.

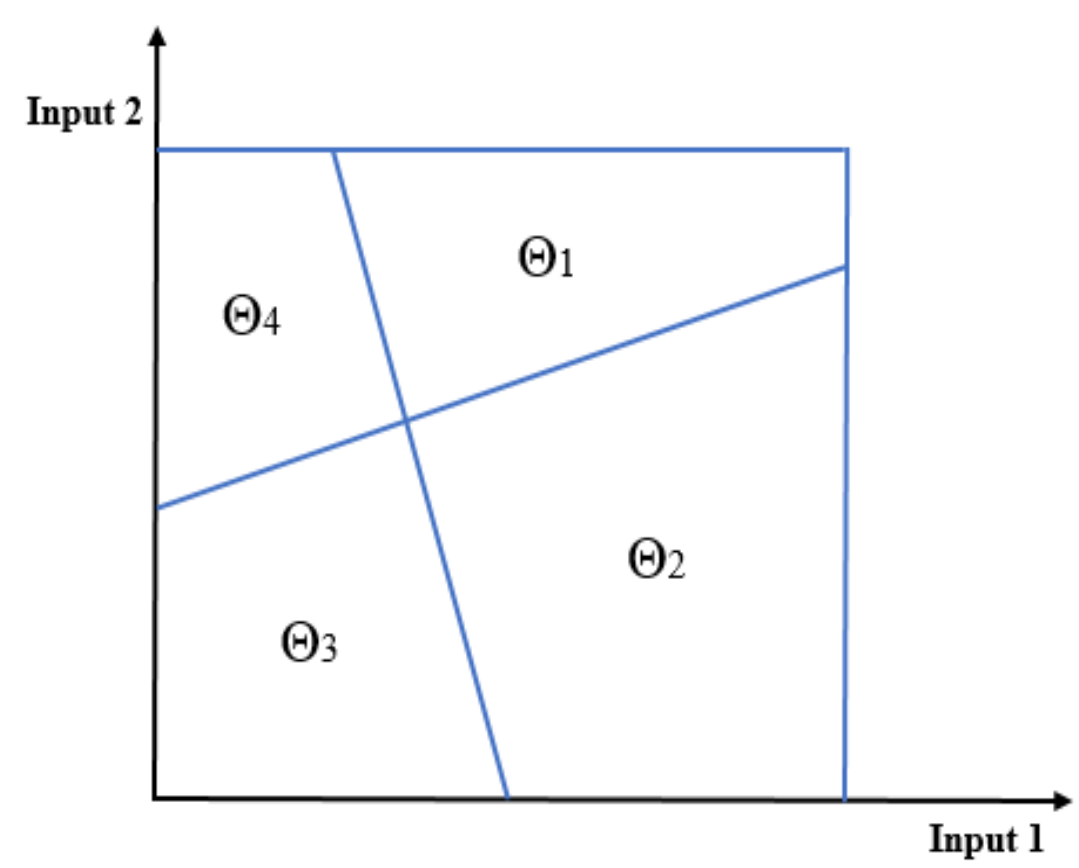

Fig. 4. The critical region in a two-dimensional space.

The OPF model introduced in Section II can be further expressed as the following nonlinear programming problem:

$$\min F = f(\boldsymbol{X}) \tag{14}$$

$$h_i(\boldsymbol{X}) = 0, \ [\lambda_i] \ \text{ for } i = 1, ..., n \tag{15}$$

$$g_j(\boldsymbol{X}) \leq 0, \ [\sigma_j] \ \text{ for } j = 1, ..., m \tag{16}$$

where $\boldsymbol{X}$=[$\boldsymbol{PG}$, $\boldsymbol{QG}$, $\boldsymbol{PF}$, $\boldsymbol{QF}$, $\boldsymbol{V}$, $\boldsymbol{\theta}$] is the vector of system control variables and state variables; $f$ is the objective function; $\boldsymbol{\lambda}$ and $\boldsymbol{\sigma}$ are the vectors of dual multipliers; $h$ represents the constraints of power flow equations; $g$ represents the inequality constraints. The KKT condition of OPF model can be formulated as follow:

$$\nabla_X L(\boldsymbol{X}, \boldsymbol{\lambda}, \boldsymbol{\sigma}) = 0 \tag{17}$$

$$\sigma_j \geq 0 \quad \text{for } j = 1, \ldots, m \tag{18}$$

$$\sigma_j g_j(\boldsymbol{X}) = 0 \quad \text{for } j = 1, \ldots, m \tag{19}$$

$$g_j(\boldsymbol{X}) \leq 0 \quad \text{for } j = 1, \ldots, m \tag{20}$$

$$h_i(\boldsymbol{X}) = 0 \quad \text{for } i = 1, \ldots, n \tag{21}$$

The set of active and inactive constraints are presented as follows:

$$\text{Active constraint (set } J\text{): } g_j(\boldsymbol{X}) = 0 \quad \text{for } j \in J \tag{22}$$

$$\text{Inactive constraint (set } J^c\text{): } g_j(\boldsymbol{X}) < 0 \quad \text{for } j \in J^c \tag{23}$$

Note that (19) includes a logical judgment, which is the key difficulty of the OPF problem. For the OPF samples that share the same active constraints, according to the *slackness complementary condition*, (19) has been slacked, and the KKT condition can be modified as follows:

$$\begin{cases} \nabla_X L(\boldsymbol{X}, \boldsymbol{\lambda}, \boldsymbol{\sigma}) = 0 \\ g_j(\boldsymbol{X}) = 0 \quad \text{for } j \in J \\ \sigma_j = 0 \quad \text{for } j \in J^c \\ h_i(\boldsymbol{X}) = 0 \quad \text{for } i = 1, \ldots, n \end{cases} \left( \begin{matrix} \sigma_j \geq 0 \quad \text{for } j \in J \\ g_j(\boldsymbol{X}) < 0 \quad \text{for } j \in J^c \end{matrix} \right) \tag{24}$$

In a specific critical region, this optimization problem is reduced to solve a set of nonlinear equations in (24). The solving process of the OPF problem is significantly improved, which is similar to the power flow calculation. The sensitivity of (24) can be easily computed by the Jacobian matrix, thus reliving the learning difficulty of the SELM network.

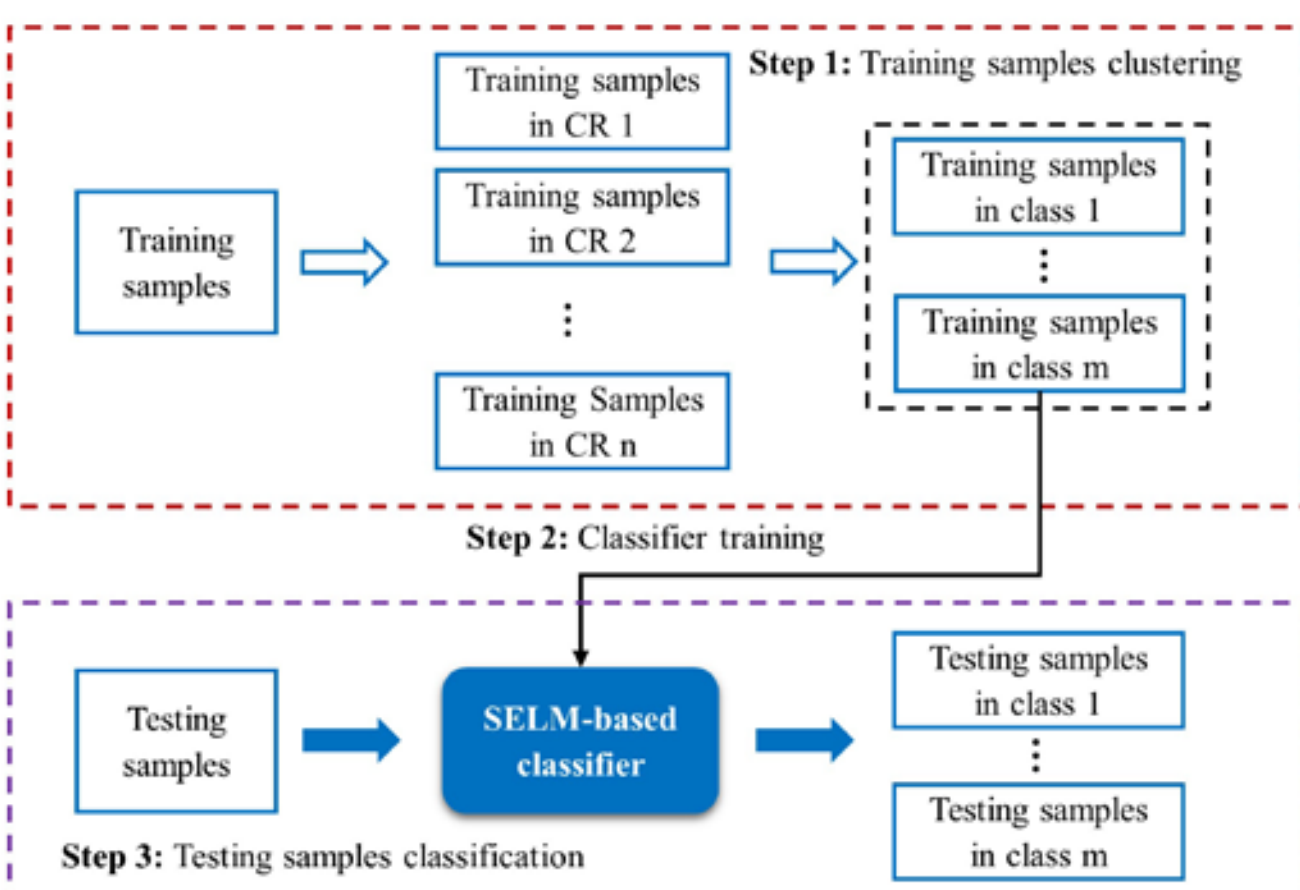

Fig. 5. Framework for sample pre-classification.

## B. Proposed Framework

There exist significant technical challenges for classification. First, the active constraints of testing samples need to be identified. Second, for the OPF problem, the combination of active constraints can be infinite. The vast number of critical regions may affect the classification precision, which is also unsuitable for SELM regression. Hence, in the proposed strategy, the critical region is obtained based on the part of active constraints. Moreover, the samples with similar active constraints are clustered together.

Based on the above discussions, the strategy for sample pre-classification is proposed in Fig. 5, which has three steps.

1) Step 1: Training samples clustering. The training samples in a specific critical region can be identified naturally based on the OPF model. To further simplify the classification, the training samples with similar active constraints are clustered into m classes.

2) Step 2: Classifier training. After clustering the training samples, the labels are obtained to train the SELM-based classifier. The input layer is designed as $PD_i, QD_i$ and the output layer is designed as the class labels. The hidden layer design is similar to Section II-C.

3) Step 3: Testing sample classification. The SELM-based classifier trained in Step 2 is used to find the class where the testing sample belongs. Then the testing samples of each class are identified.

The proposed sample pre-classification strategy further improves the learning performance by clustering samples with similar active constraints. For OPF regression, the mapping relationship between the input and output is mitigated because of the reduction of the OPF model complexity. Note that the constraints and variables of samples in a specific cluster remain the same.

## C. Flowchart of The Proposed Approach

The proposed approach provides a data-driven SELM learning framework to learn the complicated relationship between the system operating status and the OPF solution. The overall flowchart is shown in Fig. 6. The four steps are elaborated as follows:

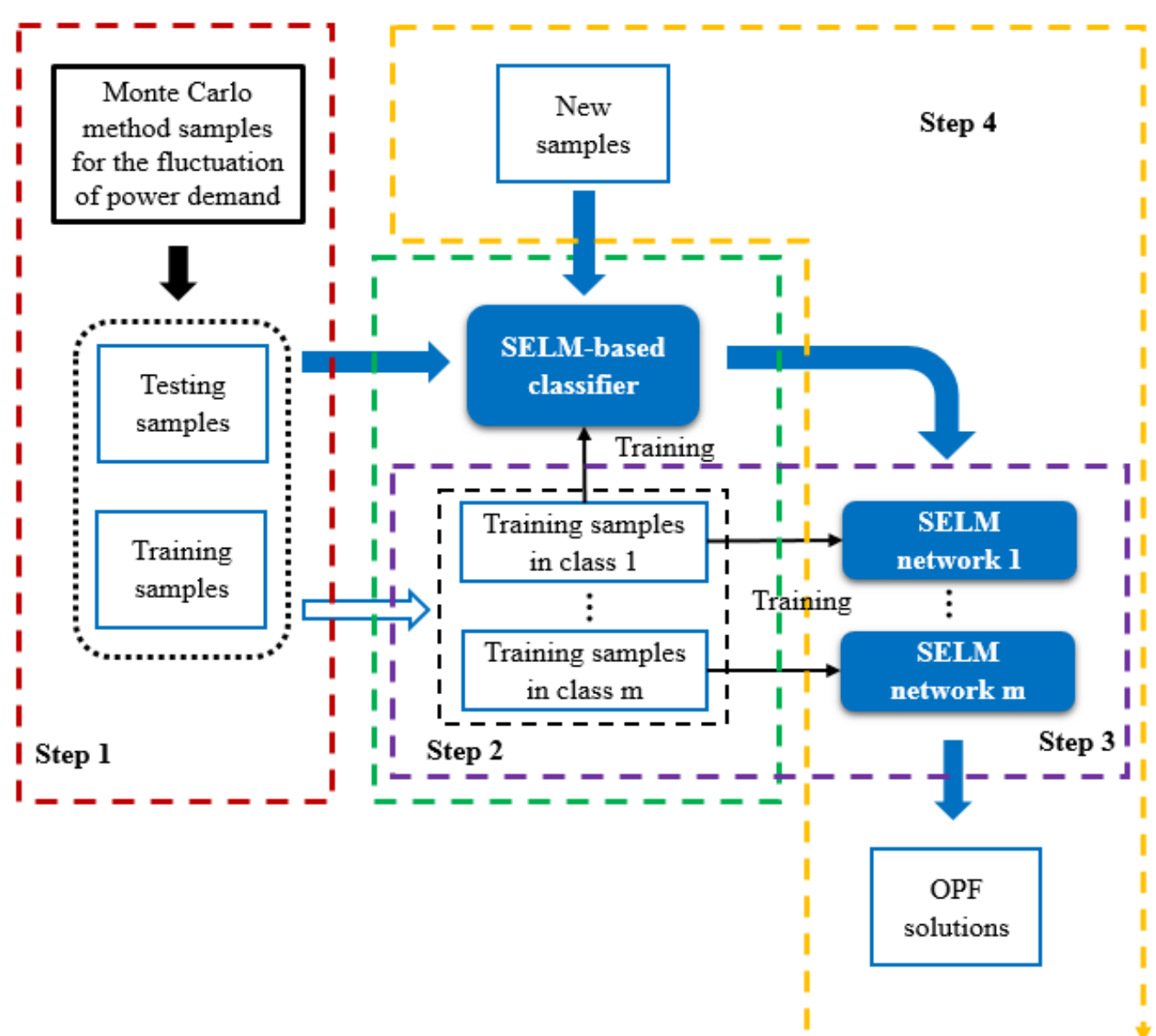

Fig. 6. Flowchart of the proposed approach.

1) Step 1: Training samples collection. For a certain power system, the Monte Carlo simulation and traditional interior-

point algorithm are used to obtain the training samples; the system power demand $PD_i, QD_i$ are regarded as the input, and the OPF solutions $PG_i, QG_i, V_i, \theta_i, PF_k, QF_k, F$ are regarded as the output.

2) Step 2: Sample pre-classification. The proposed sample pre-classification strategy is used to cluster the training samples. After clustering, their class labels are obtained to train the SELM-based classifier.

3) Step 3: SELM network training. Based on the data-driven OPF regression framework, several SELM networks are trained via the corresponding training samples, and the testing samples classified by the SELM-based classifier are used to verify the learning accuracy.

4) Step 4: Data-driven calculation. Input new $PD_i, QD_i$ obtained in practice into the trained SELM neural network. Then, OPF solutions can be calculated with high precision and fast computation.

## IV. CASE STUDY

### A. Case Setting

Numerical test cases are carried out on IEEE 39, 57, 118-bus, and Polish 2383-bus systems with the integration of renewables to validate the proposed data-driven OPF. Specifically, wind power and photovoltaic are connected to different buses. Photovoltaic generation obeys the beta distribution, and wind power obeys the Weibull distribution. The renewable penetration rate and load fluctuation rate are shown in Table I. A variety of methods is compared, as shown in Table II. The hardware and software used in the simulation are listed as follows: Intel i7-8700K CPU, 32G RAM, WINDOWS 10, and MATLAB2018b.

The accuracy index *p* is introduced to measure the accuracy of the learning performance, which is the probability of the learning error less than the threshold *thr*, as shown in (25).

$$p = P\left(\left|\hat{T} - T\right| < thr\right) \tag{25}$$

where $\hat{T}$ and $T$ is the predicted value and actual value, respectively.

For $V$ and $\theta$, the threshold *thr* is set as 0.001p.u and $0.5^0$, respectively. For *PF*, *QF*, *PG*, and *QG*, the threshold *thr* represents one percent of the average value of itself in the training data. While for the objective function value F, the threshold *thr* is chosen as one-thousandth of its average value in the training data.

### B. Evaluation of the Proposed Approach

To demonstrate the achieved benefits by the proposed approach, we compare the learning performance of M3, M4, M5, and M6 in the IEEE 39-bus system, as shown in Table III. For the sample pre-classification, the class number *m* is set as 2, and the voltage magnitude constraint is chosen to identify the critical region. The number of training and testing samples is set as 30000 and 10000, respectively. The hidden neurons *L* and the reduced hidden neurons *l* are set as 1000 and 100, respectively. Note that the number of reduced hidden neurons is chosen as 10%×*L*, a commonly used value [31]. The iterative number of single SELM is set as 10. The number of layers in the reinforcement mode is set as 2. The penalty factor C is set as $2^{-30}$, a value close to 0. According to the comparison results, the following conclusions can be drawn:

TABLE I
UNCERTAINTIES IN CASES

| Cases | Renewable penetration rate | Load fluctuation rate |
|---|---|---|
| IEEE 39 | 27.49% | 10% |
| IEEE 57 | 28.34% | 10% |
| IEEE 118 | 27.82% | 10% |
| Polish 2383 | 10.26% | 10% |

1) For M3 and M4, we find that direct learning is intractable by original SELM because the OPF model features are complicated. The proposed OPF regressive framework relieves the learning complexity of the OPF model by decomposing the learning task into three stages. The learning performance of each variable is improved significantly.

TABLE II
METHODS FOR COMPARISON

| Method | | Details | | |
|---|---|---|---|---|
| M0 | | Traditional interior point algorithm (benchmark for data-driven methods). | | |
| M1 | | A DL algorithm based on SDAE [38]. | | |
| M2 | | A DL algorithm based on SAE [39]. | | |
| | | OPF regressive framework | Reinforcement mode | Sample pre-classification |
| M3 | Original SELM [28] | × | × | × |
| M4 | Proposed method with reduced features | ○ | × | × |
| M5 | | ○ | ○ | × |
| **M6** | **Proposed method** | ○ | ○ | ○ |

[a]Symbol ○ is considering this method and × is not.

TABLE III
IMPROVEMENT OF THE PROPOSED METHOD IN IEEE 39-BUS SYSTEM

| Method | Training time (s) | Testing time (s) | Accuracy index *p* (%) | | | | | | |
|---|---|---|---|---|---|---|---|---|---|
| | | | *V* | *θ* | *PF* | *QF* | *PG* | *QG* | *F* |
| M3 | 13.53 | 2.054 | 91.74 | 99.67 | 86.19 | 74.13 | 95.13 | 74.54 | 98.76 |
| M4 | 36.14 | 4.923 | 97.25 | 99.93 | 86.24 | 74.21 | 99.18 | 91.55 | 99.94 |
| M5 | 70.86 | 11.805 | 99.00 | 99.94 | 98.17 | 92.17 | 99.60 | 94.67 | 99.94 |
| **M6** | **83.23** | **9.195** | **99.07** | **99.98** | **98.85** | **94.92** | **99.80** | **95.58** | **99.98** |

TABLE IV
COMPARISONS OF THE LEARNING PERFORMANCE

| Cases | Variables | Accuracy index $p$ (%) | | | Variables | Accuracy index $p$ (%) | | |
|---|---|---|---|---|---|---|---|---|
| | | **M6** | M1 | M2 | | **M6** | M1 | M2 |
| IEEE 39 | $V$ | **99.07** | 97.24 | 99.02 | $PF$ | **98.85** | 62.85 | 73.34 |
| | $\theta$ | **99.98** | 99.37 | 99.57 | $QF$ | **94.92** | 72.70 | 88.86 |
| IEEE 57 | $V$ | **98.56** | 89.19 | 85.46 | $PF$ | **99.64** | 90.64 | 87.80 |
| | $\theta$ | **99.99** | 99.32 | 99.03 | $QF$ | **99.58** | 98.85 | 98.54 |
| IEEE 118 | $V$ | **99.55** | 91.62 | 91.36 | $PF$ | **98.02** | 54.01 | 53.80 |
| | $\theta$ | **99.95** | 96.81 | 96.04 | $QF$ | **98.66** | 92.96 | 91.81 |
| Polish 2383 | $V$ | **97.66** | 62.06 | 59.42 | $PF$ | **98.74** | 53.73 | 52.97 |
| | $\theta$ | **99.88** | 97.09 | 95.74 | $QF$ | **97.42** | 88.56 | 87.93 |
| IEEE 39 | $PG$ | **99.80** | 97.01 | 99.23 | $F$ | **99.94** | 0.90 | 28.32 |
| | $QG$ | **95.58** | 82.59 | 89.69 | | | | |
| IEEE 57 | $PG$ | **98.97** | 79.04 | 72.68 | $F$ | **99.62** | 11.24 | 6.41 |
| | $QG$ | **99.02** | 74.80 | 65.67 | | | | |
| IEEE 118 | $PG$ | **98.76** | 80.83 | 83.01 | $F$ | **99.99** | 4.36 | 17.45 |
| | $QG$ | **98.95** | 65.84 | 65.59 | | | | |
| Polish 2383 | $PG$ | **99.21** | 82.47 | 82.13 | $F$ | **99.93** | 16.03 | 15.81 |
| | $QG$ | **98.23** | 98.16 | 98.01 | | | | |

TABLE V
COMPARISON OF TRAINING AND TESTING TIME

| Cases | Training time (s) | | | | Testing time of 10000 samples (s) | | | |
|---|---|---|---|---|---|---|---|---|
| | **M6** | M0 | M1 | M2 | **M6** | M0 | M1 | M2 |
| IEEE 39 | **63.48** | - | 4082 | 3974 | **6.425** | 663.7 | 1.458 | 1.512 |
| IEEE 57 | **85.79** | - | 15015 | 15054 | **9.556** | 877.1 | 1.1856 | 1.0296 |
| IEEE 118 | **111.8** | - | 18331 | 16458 | **10.398** | 1131 | 0.9204 | 0.9516 |
| Polish2383 | **27642** | - | 172440 | 167852 | **844.313** | 17894 | 15.507 | 15.224 |

2) For M4 and M5, thanks to the reinforcement mode, the learning performances are enhanced. Specifically, in the reinforcement mode, two supervised layers are used to make a trade-off between the accuracy and the network complexity. Note that too many layers are unnecessary as this provides negligible benefits in improving the learning accuracy.

3) Comparing M6 with M5, the learning accuracy of each variable is further improved because the complicated mapping relationship between the input and output is mitigated by the proposed sample pre-classification strategy.

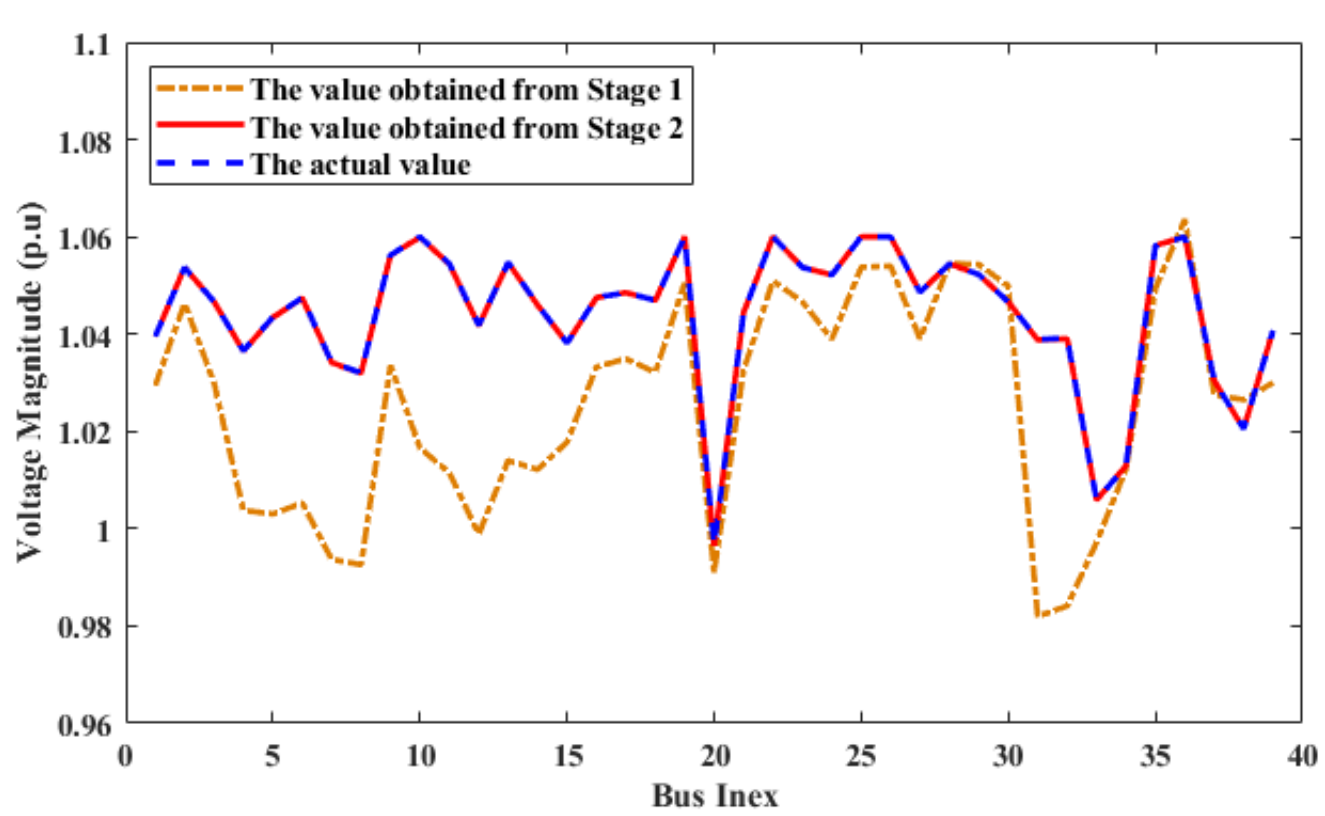

Fig. 7. Comparison of the voltage magnitude in the IEEE 39-bus system.

The error correction of the proposed three-stage SELM network is also discussed here. To demonstrate that, the voltage magnitude (learning target of Stage 2) and active power output of generator (learning target of Stage 3) are calculated by the physical power flow model leveraging the output of Stage 1 (i.e., $PF_k, QF_k$) and Stage 2 (i.e., $V_i, \theta_i$), respectively. For the voltage magnitude, the values obtained from Stage 1 and Stage 2 are compared with the actual ones, as shown in Fig. 7. Meanwhile, the error of active power output obtained from Stage 2 and Stage 3 is compared in Fig. 8.

Note that the error of voltage magnitude has been corrected in Stage 2, and Stage 3 is mainly used to correct the error of control variables and objective function value. The learning accuracy has been improved through the three stages to meet the accuracy requirement.

### *C. Comparison Results with Existing Methods*

In this subsection, the performance of M0, M1, M2, and M6 are compared, and their results are shown in Table IV and Table V. For the sample pre-classification, the class number m is set as 2, and the voltage magnitude constraint is chosen to identify the critical region. For the IEEE 39, 57, and 118-bus systems, the hyperparameters are set to be the same as Section V-B. The deep learning-based networks are set to 4 hidden layers and 400 neurons per layer. The pre-training and fine-tuning are set as 200 and 500 epochs, respectively. For the Polish 2383-bus system, the deep learning-based networks are set to 4 hidden layers and 700 neurons per layer. For our proposed approach, the number of training and testing samples is set as 100000 and 10000, respectively. The hidden neurons $L$ and the reduced hidden neurons $l$ are set as 7000 and 700, respectively. The iterative number of single SELM is set as 30.

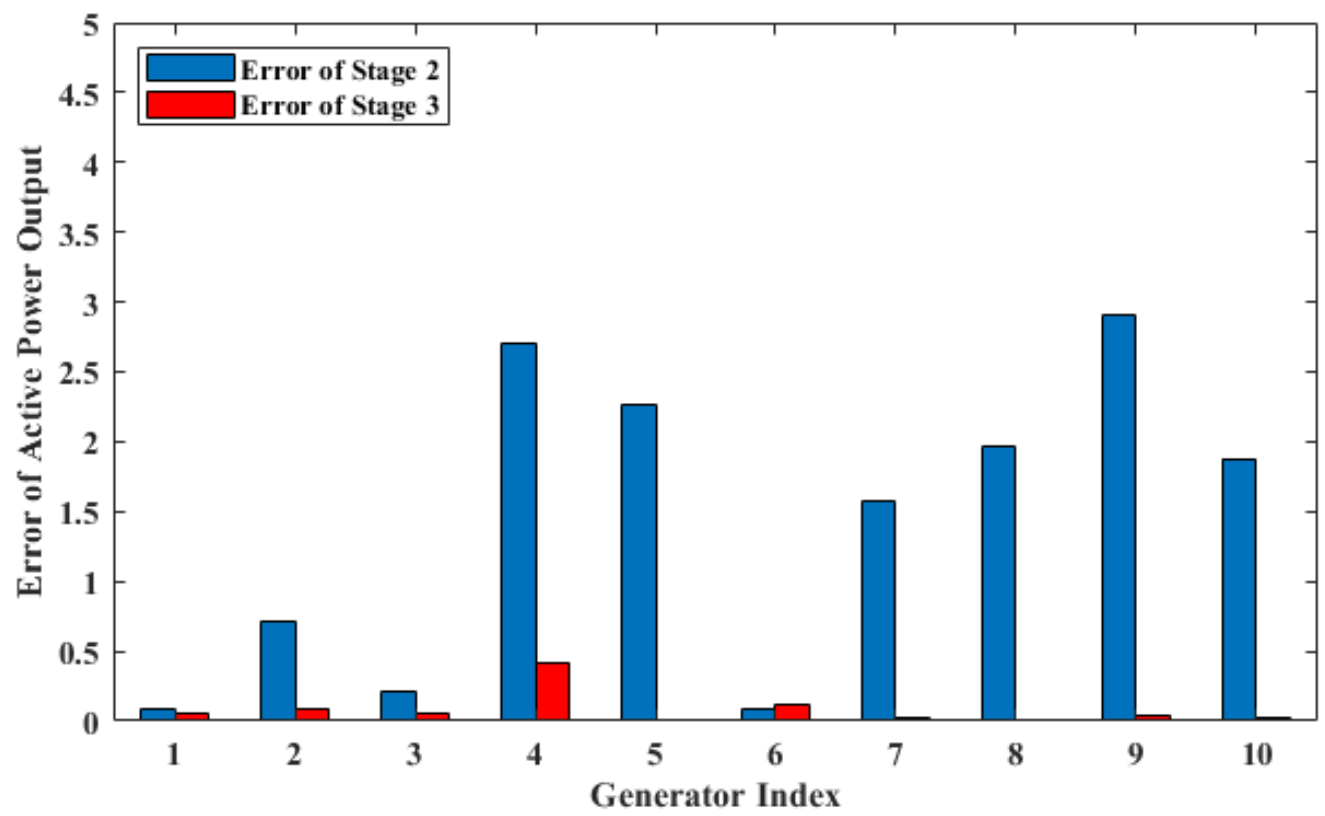

Fig. 8. Comparison of the active power output error in the IEEE 39-bus system.

Several conclusions can be drawn as follows:

1) Among all cases of OPF calculations, the proposed approach achieves the best accuracy.

2) The trained NNs in M4, M6, and M7 have acceptable testing time, and they are much faster than the time-consuming iterations in M5.

3) The training cost for the deep learning methods is much higher than the proposed approach because of the backpropagation process. Besides, deep learning methods require massive hyperparameters tuning. By contrast, with a little adjustment of hyperparameters, the proposed approach can be easily extended to different systems while retaining similar accuracy.

## V. CONCLUSION

This paper proposes a physics-informed data-driven OPF approach based on the SELM framework. The latter is further enhanced to improve its learning capability by developing a three-stage SELM learning scheme. This allows us to decompose the OPF model features and significantly reduce the learning complexity. Furthermore, a data pre-classification strategy is proposed for enhancing SELM learning performance. Compared with the deep learning algorithms, the proposed method only requires very few adjustments of the parameters and thus can be easily extended to other systems. Numerical results carried out on several IEEE and Polish benchmark systems show that the proposed method achieves the best performance as compared to other alternatives. It is also shown that the direct deep learning for the OPF may not be tractable due to the problem complexity. The future work will be on testing the developed method using realistic systems with field data.